\documentclass[useAMS,usenatbib, epsfig]{mn2e}
\include{psfig}
\newcommand{\be}{\begin{equation}}   
\newcommand{\ee}{\end{equation}}   
\newcommand{\bea}{\begin{eqnarray}}   
\newcommand{\eea}{\end{eqnarray}}

\title[Multipole invariants and non-Gaussianity]
{Multipole invariants and non-Gaussianity}
\author[Kate Land and Jo\~{a}o Magueijo]{
Kate Land and Jo\~{a}o Magueijo 
\thanks{E-mail:kate.land@imperial.ac.uk,
j.magueijo@ic.ac.uk}
\\\\
Theoretical Physics Group, Imperial College, Prince Consort Road, 
London SW7 2BZ, UK
}

\begin{document}

\date{}

\maketitle

\pagerange{\pageref{firstpage}--\pageref{lastpage}} \pubyear{2003}

\label{firstpage}

\begin{abstract}
We propose a framework for separating the information 
contained in the CMB multipoles, $a_{\ell m}$, into its algebraically 
independent components. Thus we cleanly separate information
pertaining to the power spectrum, non-Gaussianity
and preferred axis effects. The formalism builds upon the recently
proposed multipole vectors~\citep{copi,copi1,katz}, 
and we elucidate a few features regarding these 
vectors, namely their lack of statistical independence for a Gaussian
random process. In a few cases 
we explicitly relate our proposed invariants to components of the $n$-point
correlation function (power spectrum, bispectrum). We find the
invariants' distributions using a mixture of analytical and numerical methods.
We also evaluate them for the co-added WMAP first year map. 
\end{abstract}


\section{Introduction}
The remarkable  quality of the WMAP data~\citep{wmap} has 
led us to a new era in observational cosmology. Yet, various
claims for ``unexpected'' non-Gaussian signals cast a shadow on the
purity of the 
data~\citep{copi,erik2,coles,cruz,hansen,hbg,vielva,Komatsu2,park,lw,muk,erik1,copi1,teg}. 
Given that there is no plausible 
theoretical explanation for these signals, a natural worry
is that the maps are not, after all,  
fully free from systematics errors or galactic 
contamination.

Most notably several groups have reported evidence for a preferred
axis being selected by large-angle fluctuations,
either in the form of multipole planarity~\citep{teg,copi,copi1,coles}, 
North-South asymmetries in the power spectrum, three-point
function, or bispectrum~\citep{erik1,hbg,us}, as well as using
other methods~\citep{erik2,hansen}. 

In assessing these asymmetries it is important to distinguish
issues of non-Gaussianity (which should be rotationally invariant)
from those of anisotropy (existence of a preferred axis).
Apart from a subtlety in the definition of statistical 
ensemble~\citep{aniso} these issues should be clearly separated.
Unfortunately no {\it systematic} approach for extracting all the 
independent invariants under rotations
from a given set of multipoles, $a_{\ell m}$, has been proposed.
The formalism of~\cite{copi} does not produce invariants.
The invariant $n$-point correlation function~\citep{conf,fmg,joao,jooes},
on the other hand, is awkward to apply and often contains redundant
information.

In this letter we remedy this deficiency in the current formalism.

\section{Statement of the problem}\label{prob}
Multipoles are irreducible representations of $SO(3)$,
so their $2\ell+1$ degrees of freedom should split into  $2\ell-2$ invariants
and 3 rotational degrees of freedom. Ideally
we would like to process the $\{a_{\ell m}\}$ for a given mutipole
$\ell$ into the power spectrum $C_\ell$ (the Gaussian degree of
freedom), $2\ell -3$ invariant measures of non-Gaussianity, and a system
of axes (assessing isotropy). In addition one should build 3 invariants
per multipole pair, for example the
Euler angles relating the two sets of multipole axes. The latter
encode inter-scale correlations.

No one has ever accomplished this project for general $\ell$,
 but see~\cite{conf} for a Quadrupole solution.
Multipoles are equivalent to symmetric traceless tensors
of rank $\ell$. The problem is then to extract the independent 
invariant contractions of these tensors plus a system of axes --
a generalisation of the concept of eigenvalues and eigenvectors.
Typically the invariants produced by this formalism are related
to the $n-$point correlation function (bispectrum, trispectrum, etc.)
The formalism becomes very complicated very quickly, and no
systematic breakdown of independent  $n-$point correlation function
components has ever been achieved.

Alternatively one may extract from the $\{a_{\ell m}\}$
a length scale and $\ell$ independent
unit vectors (the multipole vectors)
 as proposed by~\cite{copi1}, and discussed further by~\cite{katz}. 
This approach is considerably simpler
 and unsurprisingly it has been taken further.
However we see a (correctable) 
problem with this approach. The multipole vectors are not rotationally
invariant, and so they mix up the issues of isotropy and non-Gaussinity.
We propose to correct this shortcoming by taking an appropriate number
of independent inner products between these vectors. These are the
sought-after invariants, and in some simple cases we 
relate them to the $n-$point correlation function. We will also
extract from the multipole vectors
a system of axes, encoding the multipole directional information.
It is these axis that are to be used when testing isotropy.

\section{The example of the quadrupole}\label{quadru}
We start by considering the quadrupole, which as shown
in~\cite{conf} may be written as
\be\label{quad1}
\delta T_2=Q_{ij}x^i x^j
\ee
where $Q_{ij}$ is a symmetric traceless matrix and $x^i$
are cartesian coordinates on the unit sphere. From this
matrix one may extract three eigenvectors
and two independent combinations of invariant eigenvalues $\lambda_i$.
These are essentially 
 the power
spectrum $C_2$ (related to the sum of the squares of $\lambda_i$)
and the bispectrum $B_{222}$ (related to the determinant of the matrix, or
the product of $\lambda_i$). 

In contrast, following the formalism of~\cite{katz}
one writes
\be\label{quad2}
\delta T_2=
A_2{\left(L^1_i L^2_j -{1\over 3}\delta_{ij}L^1\cdot L^2\right)}x^i x^j
\ee
where $A$ is a scale and $\{L^1_i,L^2_i\}$ are two units vectors,
encoding the information on non-Gaussianity and anisotropy.

The vectors $\{L^1_i,L^2_i\}$ are not invariant, but one
can construct an invariant by taking $X=L^1 \cdot L^2$.
As pointed out to us by~\cite{priv}, 
one may easily check that the eigenvectors of $Q$ are
\bea
V^1&=&L^1 + L^2\\
V^2&=&L^1 - L^2\\
V^3&=&L^1\times L^2
\eea
These have corresponding  eigenvalues  
$A(X+3)/6$, $A(X-3)/6$, and $-A X/3$, respectively. 
Using results in~\cite{conf}
one may therefore prove that the power spectrum and bispectrum
are:
\bea
C_2&=&\frac{A_2^2}{6}(X^2+3)\\
B_{222}&=&\frac{A_2^{3}X}{2^{2}3^{3}}(9-X^2)
\eea
so that the normalised bispectrum is
\be
I_2=\frac{B_{222}}{C_2^{\frac{3}{2}}}
=\frac{X(9-X^2)}{(X^2+3)^{3/2}}
\ee
These formulae bridge the two formalisms.

\begin{figure}
\centerline{\psfig{file=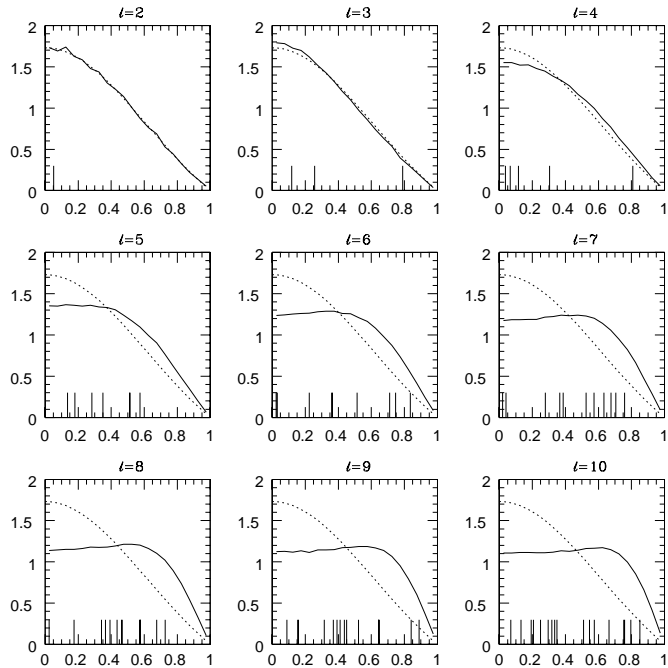,width=9cm}}
 \caption{The average distribution of the modulus
of the $2\ell-3$ dot products $X_{ij}$ 
(with the two anchor vectors chosen at random)
for multipoles $\ell=2-10$. The dotted line corresponds to
the analytical expression found for the quadrupole. We have also
plotted the invariant values for the 
WMAP first year data, with the 
anchor vectors chosen at random (short lines along the bottom
of each panel).}
\label{fig1}
\end{figure}

It was proved in~\cite{conf} that for a Gaussian process $C_2$, $I_2$ 
and the eigenvectors
are independent random variables, and that
$I_2$ is uniformly distributed in the range $[-1,1]$ (and 
$C_2$ has a $\chi^2_5$ distribution). We thus can
prove that $X$ is statistically
independent from $C_2$ (but not from $A$) and from the 
eigenvectors. By directly evaluating the Jacobian of the transformation
we find that its distribution is
\be\label{PX}
P(X)=27{1-X^2\over (X^2+3)^{5/2}}
\ee
that is, it is not uniform. We have confirmed this result with
Monte-Carlo simulations (see Fig.~\ref{fig1}, top left panel).

This elucidates an interesting feature of multipole vectors.
Even though they are algebraically independent (in the
sense that they contain no
redundant information given a concrete quadrupole realization)
they are {\it not statistically independent}.
If vectors $L^1$ and $L^2$ were statistically isotropic (which they are) 
and statistically independent, then $X$ would be uniformly
distributed. As (\ref{PX}) shows this is not the case; hence
vectors $L^1$ and $L^2$ are statistically correlated.
Specifically they prefer being orthogonal to being aligned.

\section{The general procedure}\label{general}

For a general multipole we have $\ell$ vectors and we could 
take as invariants all possible inner products between them. 
This would clearly lead to much redundant information, in contradiction
with the requirements laid down in Section~\ref{prob}. A possible
way out is to select two ``anchor'' vectors $L^1$ and $L^2$, consider their
dot product $X_{12}=L^1\cdot L^2$, and then for $3\leq i\leq \ell$
the two products $X_{1i}=L^1\cdot L^i$ and $X_{2i}=L^2\cdot L^i$.
We thus obtain $2\ell -3$ algebraically independent invariants.
In Fig.~\ref{fig1} we plot their distribution
for $2\leq \ell\leq 10$ after the order and direction ($\pm$) 
of the $\ell$ vectors has been randomised. The anchor 
vectors $L^1$ and $L^2$
are therefore selected at random, so that all $X_{ij}$ invariants for
a given multipole have the same distribution. We checked that 
the invariants $X$ are uncorrelated. As we can see this
distribution is $\ell$ dependent, and the tendency for multipole
vectors to seek orthogonal directions is less pronounced for higher
multipoles. We have also plotted these invariants as computed
for the WMAP co-added masked first year map (details in~\cite{us}).
No evidence for non-Gaussianity is found, but we defer a closer
scrutiny to a future publication.

\begin{figure}
\centerline{\psfig{file=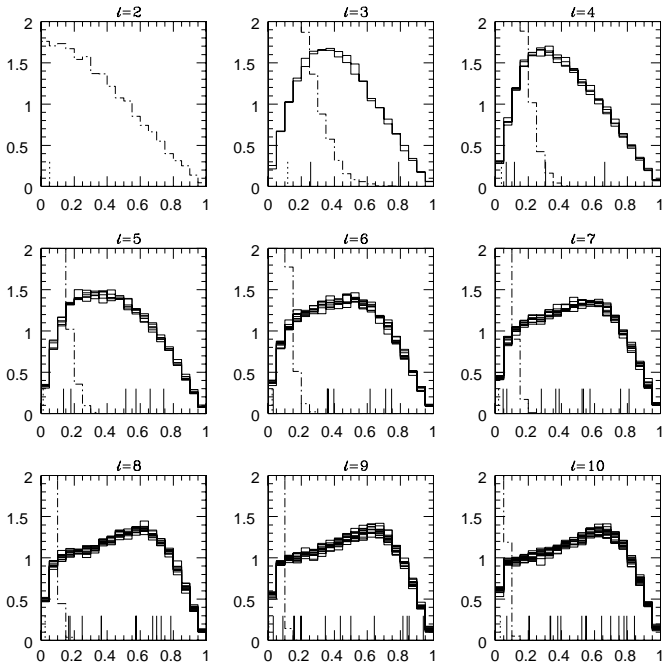,width=9cm}}
 \caption{The distribution of the modulus of the
dot products for multipoles 
$\ell=2-10$, with anchor vectors $L^1$ and $L^2$ chosen as 
the most orthogonal pair of multipole vectors. 
The solid lines are for dot products 
$X_{1i}$ and $X_{2i}$, and the dotted line shows the $X_{12}$ distribution. 
The WMAP results are also plotted (short lines along the bottom, dashed line
 the $X_{12}$ result).}
\label{fig2}
\end{figure}

The proposed procedure provides an interesting Gaussianity test.
In particular it can be easily applied to extract all the relevant
information for high~$\ell$; in contrast isolating such information
from the $n$-point correlation function is most cumbersome.
However the suggested algorithm suffers from the drawback that
the anchor vectors $L^1$ and $L^2$ are selected at random, and so
the procedure is not reproducible.

We may correct this by imposing a criterium
for their selection, for example $L^1$ and $L^2$ may be taken as the most
orthogonal among the given $\ell$ vectors. We may then form 
the ``eigenvectors'' (three orthogonal vectors):
\bea
V^1&=&L^1 + L^2\\
V^2&=&L^1 - L^2\\
V^3&=&L^1\times L^2
\eea
as the natural variables for encoding the multipole directionality.
For invariants we may take the power spectrum and the $2\ell -3$ quantities:
\bea
X_{12}&=&L^1\cdot L^2\\
X_{1i}&=&L^1\cdot L^i\\
X_{2i}&=&L^2\cdot L^i
\eea
(for $3\leq i\leq \ell$). These encode all the relevant non-Gaussian
degrees of freedom. 

The transformation from $a_{\ell m}$ into these variables is
invertible (up to discrete uncertainties related to branch choice) 
and provides a
solution to our problem, as phrased in Section~\ref{prob}.
The proposed variables for a given multipole $\ell$ are
the power spectrum $C_\ell$ (Gaussian degree of freedom), 
the $2\ell -3$ inner products $X$ (non-redundant non-Gaussian invariants),
and the orthogonal vectors $\{V^1,V^2,V^3\}$ (measures of anisotropy). 
The procedure reduces to the one found for the quadrupole when $\ell=2$.
Within this framework the inter-$\ell$ correlations are measured
by the Euler angles relating the systems of axes $\{V^1,V^2,V^3\}$
associated with each pair of multipoles. These should be uniformly distributed
for a Gaussian distribution or indeed for any theory in which
the various $\ell$ are uncorrelated.

In Fig.~\ref{fig2} we plot distributions for the $X$ invariants
with anchor vectors $L^1$ and $L^2$ defined as the most orthogonal.
We have used 12,500 realizations to make these histograms.
$P(X_{12})$ peaks around zero. The other $X$ distributions
are the same. We also plotted
the invariants for the WMAP first year data, again finding no evidence
for non-Gaussianity. We have checked that
the eigenvectors, $V^{i}$, are uniformly distributed.

Notice that the invariants $X$ cannot be 
independent variables, since their ranges of
variation are interconnected. This is to be compared with 
the lack of independence among the various vectors $L^i$,
as demonstrated in the previous Section.

\begin{figure}
\centerline{\psfig{file=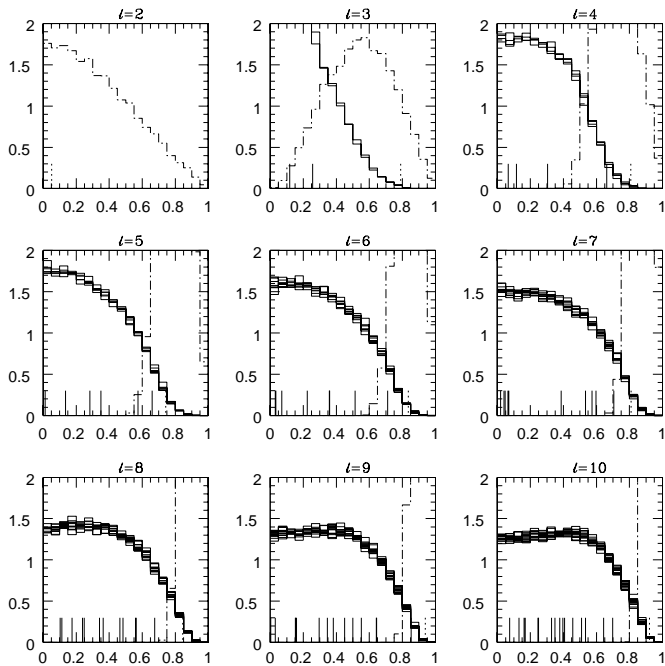,width=9cm}}
 \caption{The distribution of the modulus of dot products for multipoles 
$\ell=2-10$, with anchor vectors $L^1$ and $L^2$ chosen as 
the most aligned pair of multipole vectors. 
The solid lines are for dot products 
$X_{1i}$ and $X_{2i}$, and the dotted line shows the $X_{12}$ distribution. 
The WMAP results are also plotted (short lines along the bottom, dashed line
 the $X_{12}$ result).}
\label{fig3}
\end{figure}

Naturally we could have defined the two anchor vectors 
$L^1$ and $L^2$ in different
ways, for example the two most aligned vectors. The invariant
$X_{12}$ would then be peaked around one. We plot the counterpart
of Fig.~\ref{fig2} with this alternative definition in Fig.~\ref{fig3}.

\section{Bridging the two formalisms}
For higher multipoles, relating the proposed formalism
and the $n-$point correlation function, as done in Section~\ref{quadru}
for the quadrupole,
becomes very involved. For instance, for the octopole
the counterpart of (\ref{quad1}) and (\ref{quad2}) is:
\be
\delta T_3=Q_{ijk}x^i x^j x^k
=A_3{\left(L^1_i L^2_j L^3_k -{1\over 3}\delta_{ij}R_k\right)}x^i x^j x^k
\ee
where one should note that terms in $\{i,j,k\}$ are symmetrized.
The remainder $R_k$ is:
\be
R_k={3\over 5}L^{(1}\cdot L^{2} L^{3)}_k
\ee
Again, $Tr(Q^2)=C_3$, and therefore one can show that:
\bea
C_3&=&\frac{A_3^2}{6\cdot9}(9+7S_2+6S_2)
\eea
where
\bea
S_1&=&(L^1\cdot L^2)^2+(L^2\cdot L^3)^2+(L^3\cdot L^1)^2\\
S_2&=&(L^1\cdot L^2)\times (L^2\cdot L^3)\times (L^3\cdot L^1)
\eea
For higher multipoles
\be
Q_{i_1 ... i_\ell}
=A_\ell{\left( L^{(1}_{i_1}... L^{\ell)}_{i_\ell} -{3\over 5}
\delta_{(i_1 i_2}(L^{(1}\cdot L^{2} L^{3}_{i_3)}... L^{\ell)}_{i_\ell} 
\right)}
\ee
The components of the $n$-point correlation
function may be obtained from the various contractions
of $Q$.
They are clearly a function of the dot product of the various
multipole vectors. 
However, the formulae become progressively more 
complex to derive. With the introduction of our alternative procedure 
we suggest that these complicated formulae are not necessary.
The $X$ variables provide a better basis for describing the data.

\section{Conclusions}
Harmonic (Fourier) space is the natural arena for comparing theory and
observation for Gaussian theories. In several past studies it
was also found to be a useful ground for testing the hypothesis
of Gaussianity. In this paper we showed further that the 
degrees of freedom in the spherical
harmonic components $\{a_{\ell m}\}$  can be simply separated into
a set of algebraically independent invariants -- the power spectrum
and a set of $2\ell -3$ non-Gaussian statistics -- 
and a set of axes encoding the multipole directionality. 
The Euler angles relating sets of axes associated with pairs
of multipoles measure inter-$\ell$ correlations.
This provides an elegant answer to 
a long unsolved problem -- how to process the information
contained in a given map into its relevant non-redundant degrees
of freedom. 

Even though we have computed some of the proposed invariants for the WMAP
first year map, we defer to a future publication a more systematic
application of this framework to real data. We stress that the formalism
is easily applicable for high~$\ell$ multipoles. This is to be contrasted
with the $n$-point correlation function. Even though it is trivial
to evaluate the bispectrum at high~$\ell$~\citep{us} it becomes
very difficult to distill all the non-redundant information
contained in a given multipole in terms of sufficiently high
order components of the correlation function.
The proposed formalism is far more efficient.

\section*{Acknowledgements}
We would like to thank Dominik Schwarz, Glenn Starkman and
Jeff Weeks for discussion and help with this project.
The results in this paper have been derived using the 
HEALPix\footnote{http://www.eso.org/science/healpix/index.html}
 package~\citep{healp}, as well as the publicly available
codes described 
in~\cite{copi}\footnote{http://www.phys.cwru.edu/projects/mpvectors/}. 
This work was performed 
on COSMOS, the UK national cosmology supercomputer.

\bsp

\label{lastpage}


\begin{thebibliography}{99}

\bibitem[\protect\citeauthoryear{Bennett \& al}{2003}]{wmap}
Bennett C.L. et al., 2003a, Astrophys. J. Suppl, 148, 1
\bibitem[\protect\citeauthoryear{Coles \& al}{2003}]{coles}
Coles P. et al., 2003, astro-ph/0310252
\bibitem[\protect\citeauthoryear{Copi, Huterer \& Starkman}{2003}]{copi}
Copi C.J., Huterer D., Starkman G.D., 2003, astro-ph/0310511.
\bibitem[\protect\citeauthoryear{Cruz \& al}{2003}]{cruz}
Cruz M. et al., 2004, astro-ph/0405341
\bibitem[\protect\citeauthoryear{Eriksen \& al}{2004a}]{erik1}
Eriksen H.K. et al., 2004a, Astrophys. J, 605, 14
\bibitem[\protect\citeauthoryear{Eriksen \& al}{2004b}]{erik2}
Eriksen H.K. et al., 2004b, astro-ph/0401276
\bibitem[\protect\citeauthoryear{Ferreira \& Magueijo}{1997}]{aniso}
Ferreira P., Magueijo J., 1997, Phys.Rev, D56, 4578
\bibitem[\protect\citeauthoryear{Ferreira, Magueijo \& G\'orski}{1998}]{fmg}
Ferreira P., Magueijo J., G\'orski K., 1998, Astrophys.J, 503, 1
\bibitem[\protect\citeauthoryear{G\'orski, Hivon \& Wandelt}{1998}]{healp}
G\'orski K.M., Hivon E., Wandelt B. 1998, astro-ph/9812350
\bibitem[\protect\citeauthoryear{Hansen \& al}{2004}]{hansen}
Hansen F.K. et al., 2004, astro-ph/0402396
\bibitem[\protect\citeauthoryear{Hansen, Banday \& G\'orski}{2004}]{hbg}
Hansen F.K., Banday  A.J., G\'orski K.M.,2004,astro-ph/0404206
\bibitem[\protect\citeauthoryear{Katz \& Weeks}{2004}]{katz}
Katz G., Weeks J., 2004, astro-ph/0405631
\bibitem[\protect\citeauthoryear{Komatsu, Spergel \& Wandelt}{2003}]{Komatsu2}
Komatsu E., Spergel D.N., Wandelt B.D., 2003, astro-ph/0305189
\bibitem[\protect\citeauthoryear{Land \& Magueijo}{2004}]{us}
Land K., Magueijo J., 2004, astro-ph/0405519 
\bibitem[\protect\citeauthoryear{Larson \& Wandelt}{2004}]{lw}
Larson D.L., Wandelt B.D., 2004, astro-ph/0404037
\bibitem[\protect\citeauthoryear{Magueijo}{1995}]{conf}
Magueijo J., 1995, Phys. Lett, B342, 32. Erratum-ibid, B352, 499.
\bibitem[\protect\citeauthoryear{Magueijo}{2000}]{joao}
Magueijo J., 2000, Astrophys. J. Lett, 528, 57
\bibitem[\protect\citeauthoryear{Magueijo \& Medeiros}{2003}]{jooes}
Magueijo J., Medeiros J., 2004, MNRAS 351, L1-4 
\bibitem[\protect\citeauthoryear{Mukherjee \& Wang}{2004}]{muk}
Mukherjee P., Wang Y., 2004, astro-ph/0402602
\bibitem[\protect\citeauthoryear{Oliveira-Costa \& al}{2004}]{teg}
de Oliveira-Costa A. et al., 2004, Phys. Rev, D69, 063516
\bibitem[\protect\citeauthoryear{Park}{2003}]{park}
Park C., 2004, MNRAS, 349, 313
\bibitem[\protect\citeauthoryear{Schwarz \& al}{2004}]{copi1}
Schwarz D. et al., 2004, astro-ph/0403353
\bibitem[\protect\citeauthoryear{Starkman \& Schwarz}{2004}]{priv}
Starkman G., Schwarz D., A private communication, May 27 2004.
\bibitem[\protect\citeauthoryear{Vielva \& al}{2003}]{vielva}
Vielva P. et al., 2003, astro-ph/0310273
\end{thebibliography}
\end{document}